%% file: Staircase_journal.tex
\documentclass[journal,10pt]{IEEEtran}

\usepackage{cite}
\usepackage{epsfig}
\usepackage{epstopdf}
\usepackage{graphicx}
\usepackage{xcolor}
\usepackage{tikz}
\usetikzlibrary{arrows,positioning,shapes.geometric,circuits.logic.US}
\usepackage{pgfplots}
\usepackage{multirow}
\usepackage{upgreek}
\usepackage{amssymb}
\usepackage{amsmath}
\usepackage{cases}
\usepackage{url}
\usepackage{pifont}
\usepackage{bm}
\usepackage{amsthm}

\usepackage{tabularx}
\usepackage{booktabs}
\usepackage{filecontents}
\usepackage{subcaption}
\newcolumntype{Y}{>{\centering\arraybackslash}X}

\theoremstyle{definition} 
\theoremstyle{definition} 
\theoremstyle{definition} 
\theoremstyle{definition} 

\newcommand{\cblue}{\textcolor{blue}}
\newcommand{\cred}{\textcolor{red}}
\newcommand{\B}{\beta}
\newcommand{\Lm}{\lambda}
\usepackage[linesnumbered, ruled, vlined, noend]{algorithm2e}

\usepackage{algpseudocode}

\usepackage{array}

\newcommand{\fixme}[2]{\ifx&#2&{\leavevmode\color{red}#1}\else{\leavevmode\color{red}FIXME\{}#1{\leavevmode\color{red}\}}\footnote{{\leavevmode\color{red}#2}}\PackageWarning{Fixme}{#1: #2}\fi}

\newcommand{\newstuff}[2]{\ifx&#2&{\leavevmode\color{blue}#1}\else{\leavevmode\color{blue}FIXME\{}#1{\leavevmode\color{blue}\}}\footnote{{\leavevmode\color{blue}#2}}\PackageWarning{Newstuff}{#1: #2}\fi}

\DeclareMathOperator{\PM}{PM}

\title{Staircase codes with non-systematic polar codes}

\author{\IEEEauthorblockN{Carlo~Condo$^{1}$,~\IEEEmembership{Senior~Member,~IEEE,} Valerio~Bioglio$^{2}$,~\IEEEmembership{Member,~IEEE,}\\
		Charles~Pillet$^{2}$,~\IEEEmembership{Student Member,~IEEE,}
        Ingmar~Land$^{2}$,~\IEEEmembership{Senior~Member,~IEEE}\\}
\IEEEauthorblockA{$^{1}$Infinera Canada Inc, Ottawa\\
$^{2}$Optical Communications Technology Lab, Huawei Technologies France SASU \\
Email: ccondo@infinera.com, $\{$valerio.bioglio,charles.pillet1,ingmar.land$\}$@huawei.com}} 

\begin{document}

\maketitle
\begin{abstract}
%
In this work we propose an encoding and decoding framework for staircase codes based on non-systematic polar codes as component codes.  The staircase structure allows for efficient parallelized decoding, while the polar component codes allow to benefit from the flexible structure and efficient soft-decision decoding algorithms.  
To enhance the performance of the polar staircase codes, we concatenate the polar component codes with cyclic redundancy check (CRC) outer codes, and we add interleavers within the staircase structure that are specific to polar code properties.  The CRCs also allow to substantially reduce the decoding complexity.
Simulation results evaluate the gain brought by our proposed techniques, and analyze the dependence of the error-correction performance on code and decoder parameters.
Comparison with the state of the art on staircase polar codes shows an improvement in BER up to 0.9~dB, or considerable complexity reduction at the same BER.
 
\end{abstract}


\IEEEpeerreviewmaketitle

\section{Introduction} \label{sec:intro}

%

High throughput and low latency have been key constraints of transmission systems in optical communications, and they are becoming increasingly important in wireless communications.  This holds particularly the forward error correction (FEC) coding.  In the optical area, staircase codes with algebraic component codes and hard-decision decoding have been adopted to provide the required very bit error rate (BER) at very low decoding complexity; for improved power efficiency, soft-decoding approaches have been considered.  In the wireless area, on the other hand, power efficiency plays a dominant role, and the requirement on the bit error rate is less stringent.  

These observations suggest as a future FEC scheme using the staircase structure to maintain high throughput and low latency, and replacing the algebraic codes by other codes that are suitable for high power efficiency under efficient soft-decision decoding.  For this purpose, we propose staircase codes based on polar codes in this paper.

Staircase codes are a powerful product-like code construction first proposed in \cite{Staircase}, an incarnation of which is standardized in the ITU-T G.709.2 recommendation for optical transport networks \cite{ITUG709.2}, in the OIF 400-ZR implementation agreement for 400Gb/s coherent links \cite{OIF400ZR}%
; they have also been considered for wireless transmission \cite{StairCaseWireless}.
They can be viewed as spatially-coupled codes with a specific interleaver shape; they introduce a memory element and are usually based on systematic algebraic codes \cite{Zipper}. 
Decoder implementations of staircase codes have lower power consumption than inherently more powerful codes, e.g. low-density parity-check codes \cite{Gallager,Urbanke/Richardson}, but the decoding of component codes plays a major role in the total decoding complexity. 

Polar codes \cite{ArikanFirst} are linear block codes that can achieve the capacity of binary-input memoryless channels at infinite code length. 
In the last decade, they have attracted the interest of the academia and industry alike; thanks to continuous research on construction techniques, performance improvements, and implementation-friendly decoding algorithms, polar codes have made their way in the 3GPP 5th generation wireless systems standard, commonly known as 5G \cite{3GPP_R15}. 

Polar codes have notable advantages with respect to algebraic codes: they are inherently rate-flexible, and decoding performance, speed and complexity can be traded off among them. 
Moreover, very low latency decoding is possible in case of code rates close to $0$ or $1$ \cite{FastDec}. 
A concatenated staircase-polar scheme has been proposed in \cite{SCPOLARConcat}, using polar codes as inner codes to exploit their flexible rate instead of the Hamming code used in the 400-ZR standard.
Systematic and non-systematic polar codes alike have been considered as component codes for product codes in \cite{par_conc_sys_pol,KoikeAkinoIrregularPT,PPC,PPCjournal}, and a staircase construction with systematic polar codes has been presented in \cite{Staircase_polar, Staircase_polar2}. 
Spatial coupling of individual polar codes via information bits, frozen bits or systematic bits have been proposed in \cite{PolarSpatiallyCoupled,PolarInfoCoupled}.

This work is an extension of the preliminary ideas presented in \cite{Staircase_OFC}, where we first proposed an encoding and decoding framework for staircase codes constructed with non-systematic polar codes. 
Non-systematic polar codes are characterized by an inherently faster and simpler encoding and decoding process with respect to systematic polar codes. 
In this paper we develop in more detail the encoding structure and greatly expand the decoding framework by considering two additional component decoding algorithms.
We propose to concatenate the polar component codes with a cyclic redundancy check (CRC) outer code that encodes part of the source bits.
Two of our decoding algorithms can greatly benefit from the CRC, while the number of decoded codewords can be almost halved. 
We then introduce alternative interleavers in the staircase construction, providing a set of designs that target the intrinsic qualities of polar component codes. 
A wide simulation campaign concludes the paper, showing the gain brought by our proposed techniques, and the relationship between the error-correction performance and a wide variety of code and decoder parameters.
Comparison with the state of the art on staircase polar codes shows an improvement in BER up to $0.9$dB. 

The remainder of the paper is structured as follows. Section~\ref{sec:prel} introduces polar codes and staircase codes. 
Section~\ref{sec:NSSPC} details the encoding and decoding of the proposed construction, component decoding, CRC concatenation and decoding reduction. 
The design of ad-hoc interleavers is addressed in Section~\ref{sec:interleaver}, while simulation results and comparisons are presented in Section~\ref{sec:simres}. 
Finally, Section~\ref{sec:conc} draws the conclusions.

\section{Preliminaries} \label{sec:prel}

\subsection{Polar codes} \label{subsec:polar}

Polar codes \cite{ArikanFirst} are linear block codes of code length $N = 2^n$, information length (dimension) $K$ and rate $R=K/N$.  They are built on the basis of the polarization effect: the $N$ bit-channels are partitioned into $K$ reliable ones, used to transmit information, and $N-K$ unreliable ones, set to a known value.
The set of frozen positions are identified by the frozen set $\mathcal{F} \subset \{0,1,\ldots,N-1\}$ of size $N-K$, which may be determined via density-evolution-based methods \cite{ArikanFirst,TrifonovDesign}.

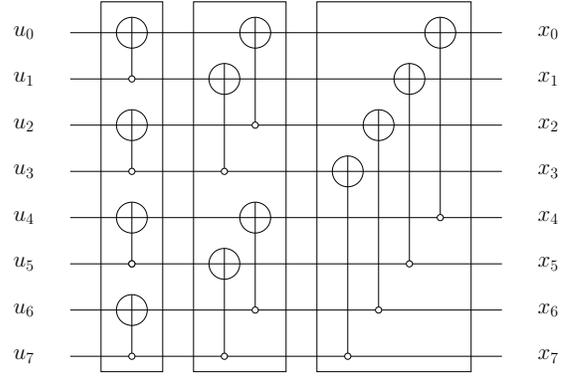
\begin{figure}
  \centering
   \scalebox{0.41}{\input{G_8.tex}}
  \caption{Factor graph for an $N=8$ polar code.}
  \label{fig:factor}
\end{figure}

Polar code encoding can be expressed as the matrix multiplication
\begin{equation}\label{eqn:enc}
\boldsymbol{x} = \boldsymbol{u}G^{\otimes n}\text{,}
\end{equation}
where $\boldsymbol{x}$ represents the codeword, $\boldsymbol{u}$ is the input vector, and the transformation matrix $G^{\otimes n}$ is the $n$-th Kronecker power of the polarizing kernel $G = \left[\begin{smallmatrix} 1&0\\ 1&1 \end{smallmatrix} \right]$. 
The input vector is constructed by assigning the $K$ information bits to the most reliable positions $j$, $j \notin \mathcal{F}$, while the $N-K$ frozen bits $j$, $j \in \mathcal{F}$, are set to $0$. 
The encoding process can be represented by a factor graph with $n$ stages of polarizing kernels, like the one shown in Figure \ref{fig:factor} for $n=3$.

Successive Cancellation (SC) decoding has been proposed in \cite{ArikanFirst} as well: it is a soft-input hard-output decoder that can be interpreted as a depth-first binary tree search with priority to the left branch. 
The logarithmic-likelihood ratio (LLR) vector $\boldsymbol{\lambda}$ at the root node is initialized as the channel observation. 
Each node at level $s$ receives a vector $\boldsymbol{\lambda}$ from its parent and computes the left $\boldsymbol{\lambda^l}$ and right $\boldsymbol{\lambda^r}$ LLR vectors to be passed to child nodes as  \cite{leroux}
\begin{align}
{\lambda}^r_i &= g_s(\lambda_i) = \lambda_{i+2^{s-1}} + (1-2\beta^{l}_{i})\lambda_{i} \text{,} \label{eq:G} \\
{\lambda}^l_i &= \tilde{f}(\lambda_{i},\lambda_{i+2^{s-1}})\text{,} \text{,} \label{eq:F}  
\end{align}
where
\begin{align*}
\tilde{f}(a,b) &\triangleq \text{min}\left(\left|a\right|,\left|b\right|\right)\text{sign}(a)\text{sign}(b) \simeq a\boxplus b  \text{,} \\
a \boxplus b &\triangleq \log\left(\frac{1+e^{a+b}}{e^a+e^b}\right) \text{.}
\end{align*}
Nodes receive the partial sums $\boldsymbol{\beta^{l}}$ and $\boldsymbol{\beta^{r}}$ from the left and right child nodes respectively, and calculate the new partial sum vector as
\begin{equation}\label{eqn:beta}
  \beta_i=\left\{
  \begin{array}{@{}ll@{}}
    \beta^{l}_{i} \oplus \beta^{r}_{i}, & \text{if}~ i \leq 2^{s-1} \\
    \beta^{r}_{i-2^{s-1}}, & \text{otherwise.}
  \end{array}\right.
\end{equation} 
where $\boldsymbol{\oplus}$ is the bitwise XOR operation. 
The partials sums of leaf nodes, which estimate the input vector $\boldsymbol{\hat{u}}$, are calculated as
\begin{equation}\label{eqn:bitestimate-sc}
\beta_{i}=\left\{
  \begin{array}{@{}ll@{}}
    0, & \text{when } \lambda_i \geq 0 \text{ } \text{or } i \in \mathcal{F}; \\
    1, & \text{otherwise.}
  \end{array}\right.
\end{equation}

\subsection{Staircase codes} \label{subsec:stair}

\begin{figure}[t!]
        \centering
        \includegraphics[clip,trim=5.2cm 3.6cm 4.8cm 1.2cm, width=0.49\textwidth]{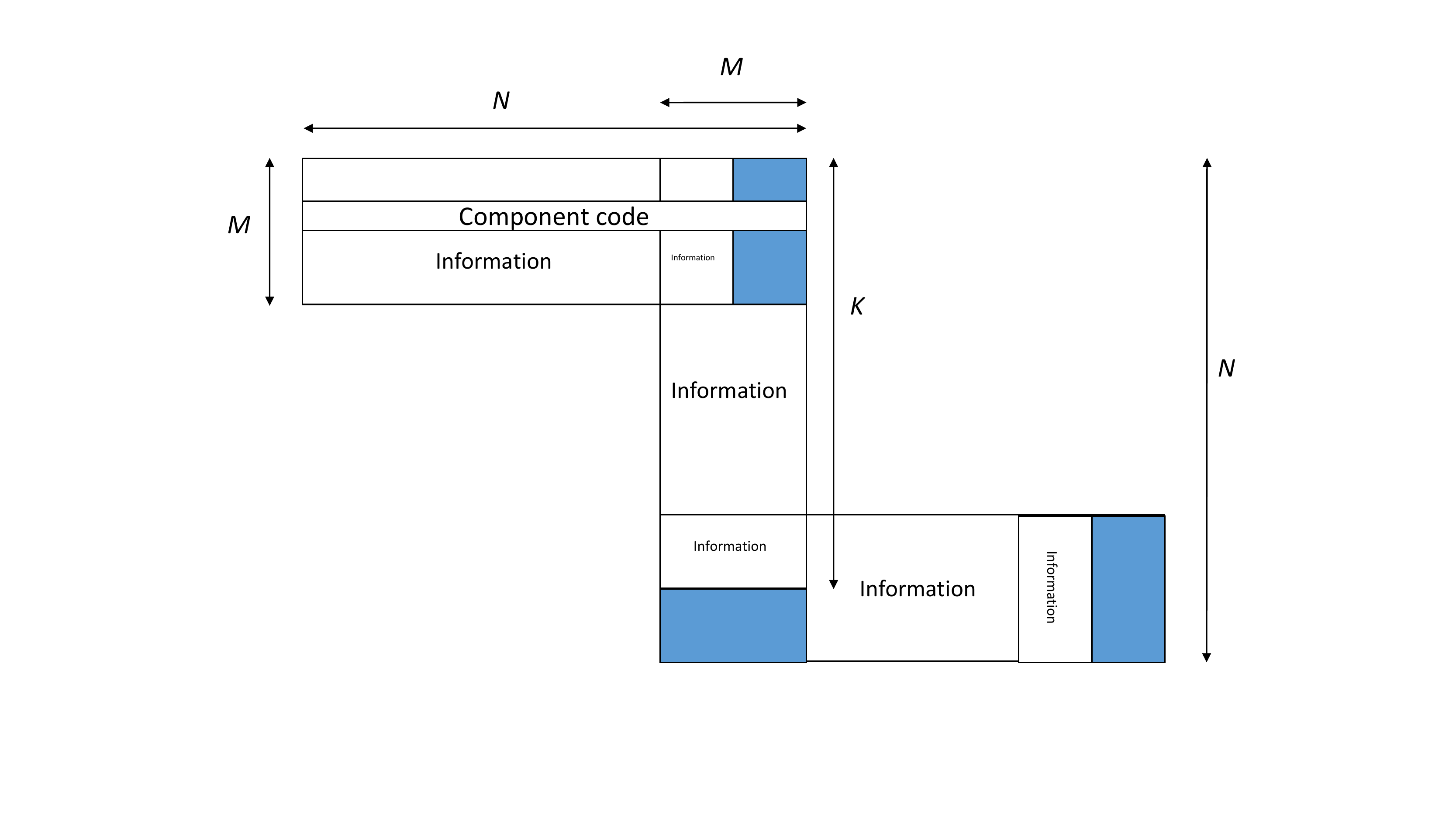}
        \caption{Staircase code construction.}
        \label{fig:staircase}
\end{figure}

Staircase codes were introduced in \cite{Staircase}, as a viable option for ITU-T standard communications. 
They merge concepts from block coding and convolutional coding; block codes are used to encode both information bits and previously encoded bits, thus introducing a memory element.
Staircase codes take their name from the graphical representation of the encoding structure, shown in Figure \ref{fig:staircase}. 
Assuming systematic component codes of length $N$ with $K$ information bits and $N-K$ parity bits (i.e. of rate $K/N$), then in each horizontal or vertical encoding step, $M \times K$ bits are encoded into $M \times (N-K)$ parity bits; note that out of these $M \times K$ bits, $M\times M$ are encoded earlier, and $M \times (K-M)$ are new source bits.  Thus the staircase code rate is $(K-M)/(N-M)$.  The straightforward construction imposes $M=N/2$.
Component codes are usually selected among algebraic codes like Bose-Chaudhuri-Hocquenghem (BCH) and Reed-Solomon, and chosen in their systematic form. 

Decoding of staircase codes can be performed as with product-like codes, through hard-decision message passing \cite{HDPC,HDSC} and soft-decision message passing \cite{SDPC,SDSC}.
Staircase codes have been shown to provide performance very close to the Shannon limit for a wide range of code rates \cite{Staircase,Staircase2}. 


Staircase codes with systematic polar component codes have been proposed in \cite{Staircase_polar, Staircase_polar2}, while preliminary work from the authors on staircase codes with non-systematic polar component codes has been published in \cite{Staircase_OFC}.

\section{Staircase Code With Non-Systematic Polar Codes} \label{sec:NSSPC}
%
%
In this section, we propose a staircase-like construction approach that makes use of non-systematic polar codes as component codes.

\begin{figure*}[t!]
    \hspace{-4mm}
    \begin{minipage}{0.5\linewidth}
       \centering
       \vspace{-0.5cm}
        \includegraphics[width=\linewidth]{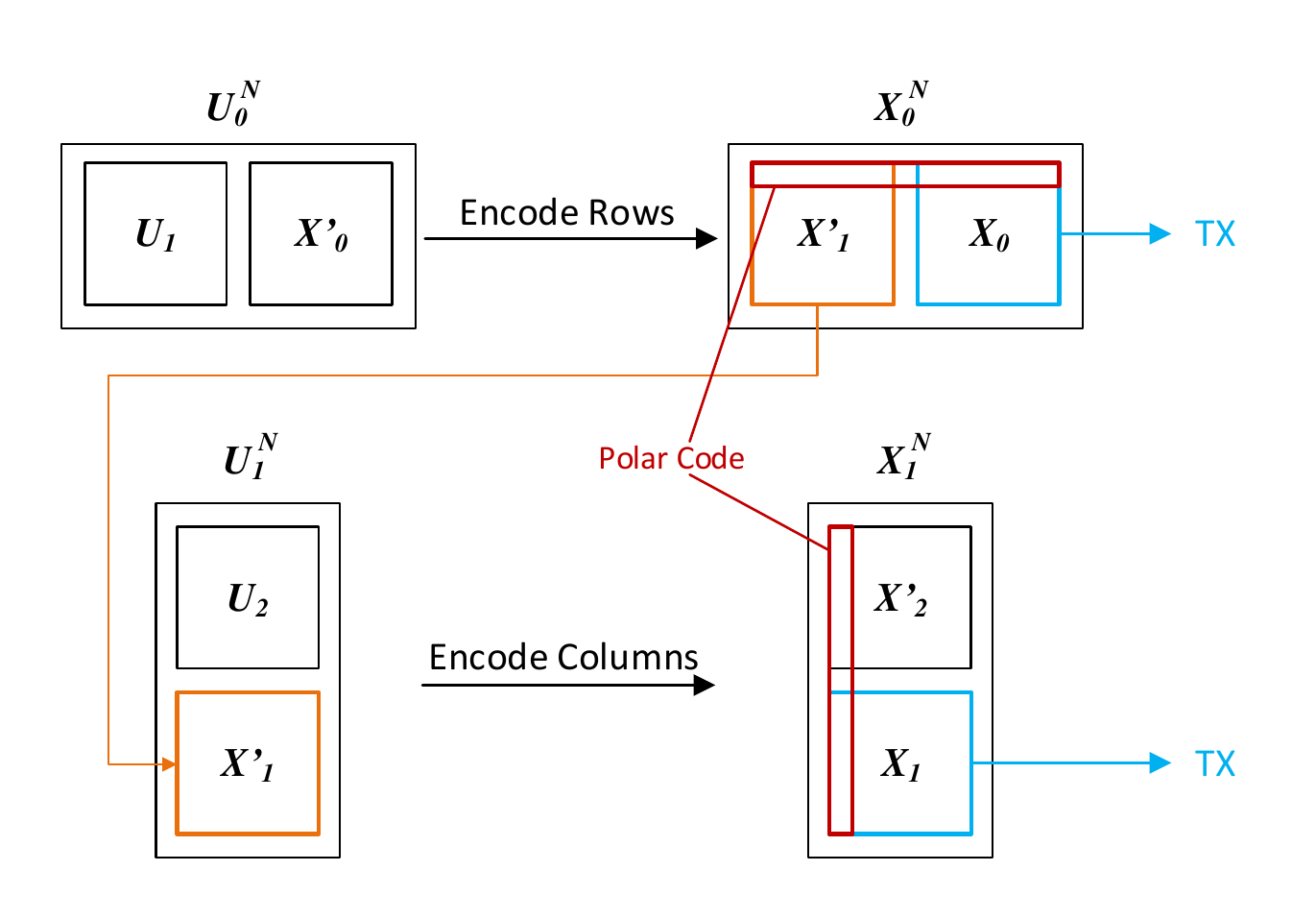}
        (a)
        \label{fig:ENC}
    \end{minipage}
    \begin{minipage}{0.5\linewidth}
            \centering
            \vspace{-1.5cm}
        \includegraphics[width=1.1\linewidth]{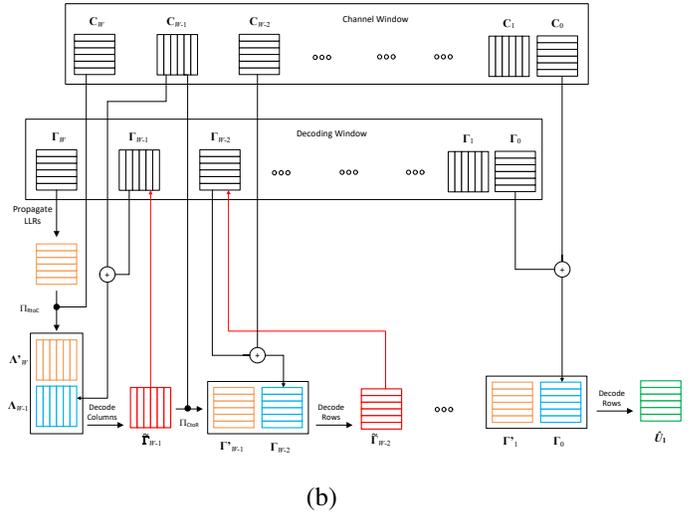}
        (b)
        \label{fig:DEC}
    \end{minipage}
    \caption{Staircase polar code encoding (a) and decoding (b) process.}
    \label{fig:ENCDEC}
\end{figure*}

%

\subsection{Construction and Encoding} \label{subsec:enc}

As explained in Section \ref{sec:prel}, in staircase-like constructions a part of the information vector (before encoding) of each component codeword is composed of previously encoded or transmitted bits.

In the most straightforward staircase construction, the $N$-bit input vector $\boldsymbol{u}$ to the polar encoder is composed of a first half $\boldsymbol{u}_\mathsf{new}$, where information and frozen bits are placed, and a second half $\boldsymbol{x'}$ that is composed of previously encoded bits, i.e., $\boldsymbol{u} = [ \boldsymbol{u}_\mathsf{new} \; \boldsymbol{x'} ]$.
Consequently, the specific $\mathcal{F}_{SC}$ used for staircase component polar codes has frozen bits in the $N - K$ least reliable positions among the leftmost $N/2$ bit positions. 

Let us define an input matrix $\boldsymbol{U}$ composed of $M=N/2$ row or column vectors $\boldsymbol{u}$ of size $N$. 
To initialize the encoding and transmission process, let us build an input matrix $\boldsymbol{U}^N_0$, where in every row $0\le i<N/2$ each bit position $0\le j \le N/2$ is an information or a frozen bit according to $\mathcal{F}_{SC}$, while all bit positions $j\ge N/2$ are set to $0$. 
Given the entries $u_{ij}$ of $\boldsymbol{U}^N_0$, we identify as $\boldsymbol{U}_1$ the matrix composed of all the entries for which $0\le j <N/2$, and as $\boldsymbol{X}_0'$ the one composed of all entries for which $N/2\le j <N$.
Each row of $\boldsymbol{U}^N_0$ is encoded according matrix $G^{\otimes n}$, resulting in the encoded matrix $\boldsymbol{X}_0^N$; as in the case of $\boldsymbol{U}^N_0$, we label as $\boldsymbol{X}_0$ the rightmost $N/2\times N/2$ matrix, and the leftmost half as $\boldsymbol{X}_1'$. 
The transmitted frame is constituted by $\boldsymbol{X}_0$, and it is discarded by the encoder. 

The columns of $\boldsymbol{X}_1'$ are instead appended to a set of column input vectors, $\boldsymbol{U}_2$, thus composing the input matrix $\boldsymbol{U}^N_1$. 
The columns of $\boldsymbol{U}^N_1$ are then encoded through $G^{\otimes n}$, resulting in the encoded matrix $\boldsymbol{X}_1^N$, within which we identify $\boldsymbol{X}_1$, composed of entries $x_{ij}$ with $N/2\le i < N$, and $X_2'$, for which $0\le i < N/2$.
$\boldsymbol{X}_1$, having been encoded by both row encoding and column encoding, is transmitted, and $\boldsymbol{X}_2'$ is interleaved and composes part of the next input matrix. 
The encoding process is repeated alternating row and column encoding, and is portrayed in Fig. \ref{fig:ENCDEC}a. 
The code structure can be generalized with $\boldsymbol{U}$ and $\boldsymbol{X}'$ being of different sizes.

\subsection{Decoding} \label{subsec:dec}

%
%
%


The decoding of staircase-like codes is based on a decoding window, i.e. a data structure that stores information relative to the latest received frame and a number of past ones. 
The decoding window is successively updated through the output of component decoders, and every time a new frame is received, one of the frames in the window is replaced, usually the oldest. 
In general, the type of information stored in the decoding window depends on both the component decoder and the staircase decoder itself. 
Given that most polar code decoding algorithms require soft input information, and that soft message passing improves the effectiveness of staircase-like code decoding, we assume that component codes are decoded with soft-in soft-out algorithms, and that soft information is represented as LLRs.

With reference to Figure~\ref{fig:ENCDEC}(b),
the decoding window is constituted of $W+1$ matrices $\boldsymbol{C}_k$, $0 \le k \le W$, each of them storing one frame of LLRs coming from the channel, and $W+1$ matrices $\boldsymbol{\Gamma}_k$, $0 \le k \le W$, where updated (extrinsic) soft information is stored;  $\boldsymbol{C}_k$ and $\boldsymbol{\Gamma}_k$ represent LLRs corresponding to the transmitted block $\boldsymbol{X}_k$.
While the decoding window can be easily implemented as a circular buffer, we can consider the last received frame being stored in $\boldsymbol{C}_W$ and the oldest frame being that in $\boldsymbol{C}_0$.  Within the decoding window, the decoder proceeds from the newest block (received last) to the oldest blocks.
In the following we explicate decoding for the newest block; decoding of the other blocks follows the same scheme.  

Assume that the last two encoding steps were the following: 
%
%
%

\noindent
after column encoding
\begin{equation*}
\begin{bmatrix}
\boldsymbol{U}_W \\ \boldsymbol{X}'_{W-1}
\end{bmatrix}  
\longrightarrow
\begin{bmatrix}
\boldsymbol{X}_W' \\ \boldsymbol{X}_{W-1}
\end{bmatrix}
\end{equation*}
$\boldsymbol{X}_{W-1}$ is transmitted; the block $\boldsymbol{X}_W'$ is then used in the row encoding
\begin{equation*}
\begin{bmatrix}
\boldsymbol{U}_{W+1} & \boldsymbol{X}'_{W}
\end{bmatrix}
\longrightarrow
\begin{bmatrix}
\boldsymbol{X}'_{W+1} & \boldsymbol{X}_{W}
\end{bmatrix} ,
\end{equation*}
after which $\boldsymbol{X}_{W}$ is transmitted.
At the receiver side, upon reception of the noisy observations of $\boldsymbol{X}_{W}$, the corresponding channel LLRs $\boldsymbol{C}_W$ are stored; the extrinsic soft information matrix $\boldsymbol{\Gamma}_W$ is initialized as zero.  The soft information $\boldsymbol{C}_k$ and $\boldsymbol{\Gamma}_k$, $k < W$, is available in the memory.

The first step of the decoding iteration aims at decoding 
$\bigl[ \begin{smallmatrix} \boldsymbol{X}_W'  \\ \boldsymbol{X}_{W-1} \end{smallmatrix} \bigr]$.
For decoding we require soft information about the blocks $\boldsymbol{X}_W'$ and $\boldsymbol{X}_{W-1}$.
The LLRs in the blocks $\boldsymbol{C}_W$ and $\boldsymbol{\Gamma}_W$ are relative to the encoded matrix $\boldsymbol{X}_W$, and 
$\boldsymbol{C}_{W-1}$ and $\boldsymbol{\Gamma}_{W-1}$ are relative instead to $\boldsymbol{X}_{W-1}$. 
The posterior LLRs for the blocks $\boldsymbol{X}_{W-1}$ and $\boldsymbol{X}_{W}$ are given by 
$\boldsymbol{\Lambda}_W = \boldsymbol{C}_W + \boldsymbol{\Gamma}_W$ and 
$\boldsymbol{\Lambda}_{W-1} = \boldsymbol{C}_{W-1} + \boldsymbol{\Gamma}_{W-1}$, respectively.
Now from the soft-information $\boldsymbol{\Lambda}_W$ relative to $\boldsymbol{X}_W$, we need to compute soft information $\boldsymbol{\Lambda}'_W$ relative to $\boldsymbol{X}'_W$.

By construction, we have $\boldsymbol{X}_W = \boldsymbol{X}_W' \cdot G^{\otimes (n-1)}$.  As the polar transform $G^{\otimes (n-1)}$ is an involution, i.e., identical to its inverse, we also have $\boldsymbol{X}_W' = \boldsymbol{X}_W \cdot G^{\otimes (n-1)}$, which represents the encoder inverse. Thus every row of $\boldsymbol{X}_W'$ can be computed by multiplying the corresponding row of $\boldsymbol{X}_W$ with the transformation matrix $G^{\otimes (n-1)}$, an operation commonly represented by a tanner graph constituted of stages of XORs, see Fig.~\ref{fig:factor}.  
To propagate soft information, we use the same Tanner graph structure, substituting the $\tilde{f}$ operator for the XORs, as expressed in (\ref{eq:F}).
 
The LLR propagation is concatenated with an interleaver $\Pi_{RtoC}$, that rotates the obtained $\boldsymbol{\Lambda}'_W$ so that they can be prepended to $\boldsymbol{\Lambda}_{W-1}$. 
Each column of $\boldsymbol{\Lambda}$ is used as an input to the component decoder, which returns extrinsic values; these are collected into the matrices $\tilde{\boldsymbol{\Gamma}}'_{W}$ and $\tilde{\boldsymbol{\Gamma}}_{W-1}$.  The block $\tilde{\boldsymbol{\Gamma}}_{W-1}$ is used to update the memory and also for decoding in the following decoding step.  The block $\tilde{\boldsymbol{\Gamma}}'_{W}$ may be processed in the reverse way as above to determine $\tilde{\boldsymbol{\Gamma}}_{W}$, which is used to update the memory; this step is not shown in Fig.~\ref{fig:ENCDEC}(b), and it may be omitted in the decoding algorithm to save complexity. 

In the next decoding step $\boldsymbol{C}_{W-1}$, $\boldsymbol{\Gamma}_{W-1}$ and $\boldsymbol{C}_{W-2}$, $\boldsymbol{\Gamma}_{W-2}$ have to be considered. 
The process is repeated alternating row and column decoding, until $\boldsymbol{C}'_1$, $\boldsymbol{\Gamma}'_{1}$ are decoded with $\boldsymbol{C}_0$, $\boldsymbol{\Gamma}_{0}$. 
The output of the decoding iteration is then the estimated $\boldsymbol{\hat{U}}_1$ hard decision matrix. 

Note that for our approach soft information about $\boldsymbol{X}_W$ needs to be converted into soft information about $\boldsymbol{X}_W'$. This is possible whenever there is a one-to-one correspondence between $\boldsymbol{X}_W$ and $\boldsymbol{X}_W'$. As generalisation of the method above using a block of length $N/2$, any shortening pattern of polar codes \cite{Bioglio:punct/short:WCNC2017} may be used to identify $\boldsymbol{X}_W'$ in $\boldsymbol{U}_W^N$.

\subsection{Component Decoding} \label{subsec:compdec}

Each polar codeword within the staircase structure can in principle be decoded according any polar code decoding algorithm. 
However, the need for soft information exchange detailed in Section \ref{subsec:dec} imposes that such algorithms accept soft values as input and return soft values as output.

In this section, we detail various soft-in soft-out polar decoding algorithms that can be used effectively for component code decoding in the proposed construction. 



\subsubsection{Soft-SCL decoding} \label{subsubsec:softSCL}

In \cite{PPCjournal} we proposed a method to obtain soft information using the successive-cancellation-list decoding algorithm (SCL) \cite{talSCL}, in particular its LLR-based formulation \cite{balatsoukas}, and apply it to product polar codes. 
This technique can be used for a staircase construction as well. 

In SCL, to each of the $L$ candidate paths is assigned a path metric $\PM$, updated after each bit estimation $\hat{u}_{i}$ as
\begin{align}
\PM_{{i}} = \begin{cases}
    \PM_{{i-1}} + |\lambda_{i}| \text{,} & \text{if } \hat{u}_{i} \neq \text{HD}(\lambda_{i})\text{,}\\
    \PM_{{i-1}} \text{,} & \text{otherwise,}
  \end{cases} \label{eq7}
\end{align}
where $\lambda_{i}$ is the LLR associated to $\hat{u}_{i}$, $\PM_0 = 0$, and $\text{HD}(\lambda_{i}) = 0$ if $\lambda_{i}\ge 0$, and $1$ otherwise.
At the end of the SCL decoding, we take the $L$ estimated input vectors $\hat{u}^{0},\dots,\hat{u}^{L-1}$, having path metrics $M_0,\dots,M_{L-1}$, and re-encode them obtaining the estimated codewords $\hat{x}^{0},\dots,\hat{x}^{L-1}$. 
Extrinsic soft information $\gamma_i$ associated to codeword bit $\hat{x}_i$ is then calculated as 
\begin{equation} \label{eq:soft}
\gamma_i=\alpha_E\left(\alpha_B\left(\min_{\hat{x}^l_i = 1}(M_l)-\min_{\hat{x}^l_i = 0}(M_l)\right) - \lambda^{in}_i\right)~,
\end{equation}
where $\lambda^{in}_i$ is the $i^{th}$ LLR input to the decoder, and $\alpha_E$ and $\alpha_B$ are scaling factors $\le1$.
In case the $L$ codewords have the same value for a given bit $\hat{x}_i$, i.e. $\{ l=0,\dots,L-1 \, s.t. \, \hat{x}^l_i = a\} = \emptyset$, $\gamma_i$ is computed as 
\begin{equation} \label{eq:soft_nocandidate}
|\gamma_i|=\alpha_E\left(\alpha_B\left(\sum_{j=k_{min}}^{k_{max}}|\dot{\lambda}^{in}_j|\right) - \lambda^{in}_i\right)~,
\end{equation} 
where $\boldsymbol{\dot{\lambda}}^{in}$ is vector $\boldsymbol{\lambda}^{in}$ sorted in ascending order of magnitude, and $k_{min}$ and $k_{max}$ are two indices devised via simulation. 
The sign of $\gamma_i$ is inferred according to $\hat{x}_i$.

\subsubsection{SCAN} \label{subsubsec:SCAN}

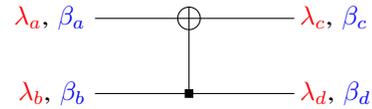
\begin{figure}[t!]
	\centering
	\input{factor-graph-2-soft}
	\caption{Factor graph with soft messages $\Lm$ and $\B$ for $N = 2$ polar code.}
	\label{fig:scan2}
\end{figure}

In \cite{SCANfirst} the authors propose an iterative decoder that follows the SC schedule, called soft cancellation (SCAN). 
Figure~\ref{fig:scan2} shows the message passing criterion on which the decoding is based, where $\Lm_{s}^{(i)}$ and $\B_{s}^{(i)}$ respectively identify the left- and right-propagating information at level $0 \leq i < N$ and stage $0\leq s < n+1$ of the factor graph (Figure~\ref{fig:factor}). 
$\Lm_{n}^{(i)}$ is initialized according to the received vector $\mathbf{y}$, while $\B_{0}^{(i)}$, that carries information about the estimated vector $\mathbf{\hat{u}}$, is set to 
\begin{equation}\label{eq:Binit}
\B_{0}^{(i)} =
  \begin{cases}
    \infty \text{,} & \text{if } i \in \mathcal{F}\text{,}\\
    0 \text{,} & \text{otherwise.}
  \end{cases}
\end{equation}
The remaining messages are initialized to 0. 
Both $\B_{0}$ and $\Lm_{n}$ are not updated through the iterations, and maintain their initial value. 
Based on the polar kernel in Figure~\ref{fig:scan2}, the SCAN message update rules are defined as
\begin{align*}
	\cred{\Lm_a} = \tilde{f}\left(\cred{\Lm_c},\cred{\Lm_d} + \cblue{\B_b}\right)~,\\
	\cred{\Lm_b} = \cred{\Lm_d} + \tilde{f}\left(\cred{\Lm_c}, \cblue{\B_a}\right)~,\\
	\cblue{\B_c} = \tilde{f}\left(\cblue{\B_a}, \cred{\Lm_d} + \cblue{\B_b}\right)~,\\
	\cblue{\B_d} = \cblue{\B_b}+\tilde{f}\left(\cblue{\B_a},  \cred{\Lm_c}\right)~.	
\end{align*}
SCAN terminates after a set number of iterations $T$. 
The estimated vector can be obtained through hard decision on stage-0 soft information:
\begin{equation}
\hat{u}_i =
  \begin{cases}
    0 \text{,} & \text{if } \Lm_{0}^{(i)}+\B_{0}^{(i)} > 0 \text{,}\\ 
    1 \text{,} & \text{otherwise.}
  \end{cases} \label{eq:SCANhard}
\end{equation}
The soft decoding output corresponds instead to $\B_n$ multiplied by a scaling factor $\alpha_E$.


\subsubsection{SCANL decoding}\label{subsubsec:SCANL}

SCAN list (SCANL) decoding has been proposed in \cite{SCANL}.
It relies on $L$ permutations of the polar code factor graph, each of which is decoded through SCAN.
All SCAN decoders return a path metric $PM$ that is used to select the most likely output:
\begin{align}\label{eq:metricSCANL}
	PM = PM - \Lm_0^{(i)} \text{ if } p(i) \in \mathcal{F}~,
\end{align}
where $p(i)$ is the permuted position of bit $i$ according to the factor graph permutation. 
As in SCL decoding, the candidate with the smallest $PM$ is chosen.
However, while in SCL decoding $PM$ increases every time the sign of an LLR does not agree with the estimated bit value, SCANL updates $PM$ only at frozen bits, either increasing or decreasing it. 
As with SCAN, the soft decoding output is the $\B_n$ set of the selected candidate, scaled by a factor $\alpha_E$.

\subsection{Decoding reduction} \label{subsec:decred}

The proposed staircase decoding schedule, similar to other sliding-window based soft decoding approaches, decodes all component codes in the window at each iteration. 
However, if it is possible to identify decoding instances that are not necessary, the number of component decoding can be reduced. 
This can have a positive impact on decoding latency and on decoder implementation complexity, as long as any error-correction performance degradation is within acceptable parameters.

We propose here to concatenate each component code source vector with a CRC, that encodes only the $K-N/2$ information bits that do not belong to $\boldsymbol{X}'$. 
Let us consider the decoding of a component code composed of $N/2$ bits of $\boldsymbol{X}'_i$ and of $N/2$ bits of $\boldsymbol{X}_{i-1}$. 
The component decoding process returns a set of soft values relative to an estimated source vector $\boldsymbol{\hat{u}}$, constructed of estimated bits belonging to $\boldsymbol{\hat{U}}_i$ and $\boldsymbol{\hat{X}}'_{i-1}$: if the CRC passes, we assume the relative row of $\boldsymbol{\hat{U}}_i$ to be correctly estimated, while no assumptions on $\boldsymbol{\hat{X}}'_{i-1}$ are made. 
We keep track of which half component codes in each frame have passed the CRC: in the following iteration, they are not decoded at all when used as $\boldsymbol{X}'$, while their $\boldsymbol{\Lambda}$ is not updated when used as $\boldsymbol{X}$.
To ensure soft value propagation throughout the decoding window, the decoding of a component codeword is forced in case it was skipped at the previous iteration. 
The soft encoding described in Section \ref{subsec:dec} is used to obtain the $\boldsymbol{\Lambda}'$ of skipped component codes.

The presence of a CRC can be beneficial to component decoders as well.
The concatenation of a polar code with a CRC has been successful in substantially improving the performance of SCL \cite{CA_SCL}: the CRC is used to select the decoder output among the $L$ available.
Candidate selection through CRC concatenation in SCANL, while useful, has instead proven to be less beneficial \cite{SCANL}, due to the lower degree of diversity in decoding approaches based on graph permutations. 
Finally, SCAN does not rely on a list of candidates, and CRC concatenation cannot be exploited straightforwardly.

The effect of CRC concatenation and of the proposed decoding reduction approach on the decoding complexity and on the error-correction performance is evaluated in Section \ref{sec:simres}.

\begin{figure*}[t!]
	\centering
	\vspace{-140pt}
	\includegraphics[width=1.1\linewidth]{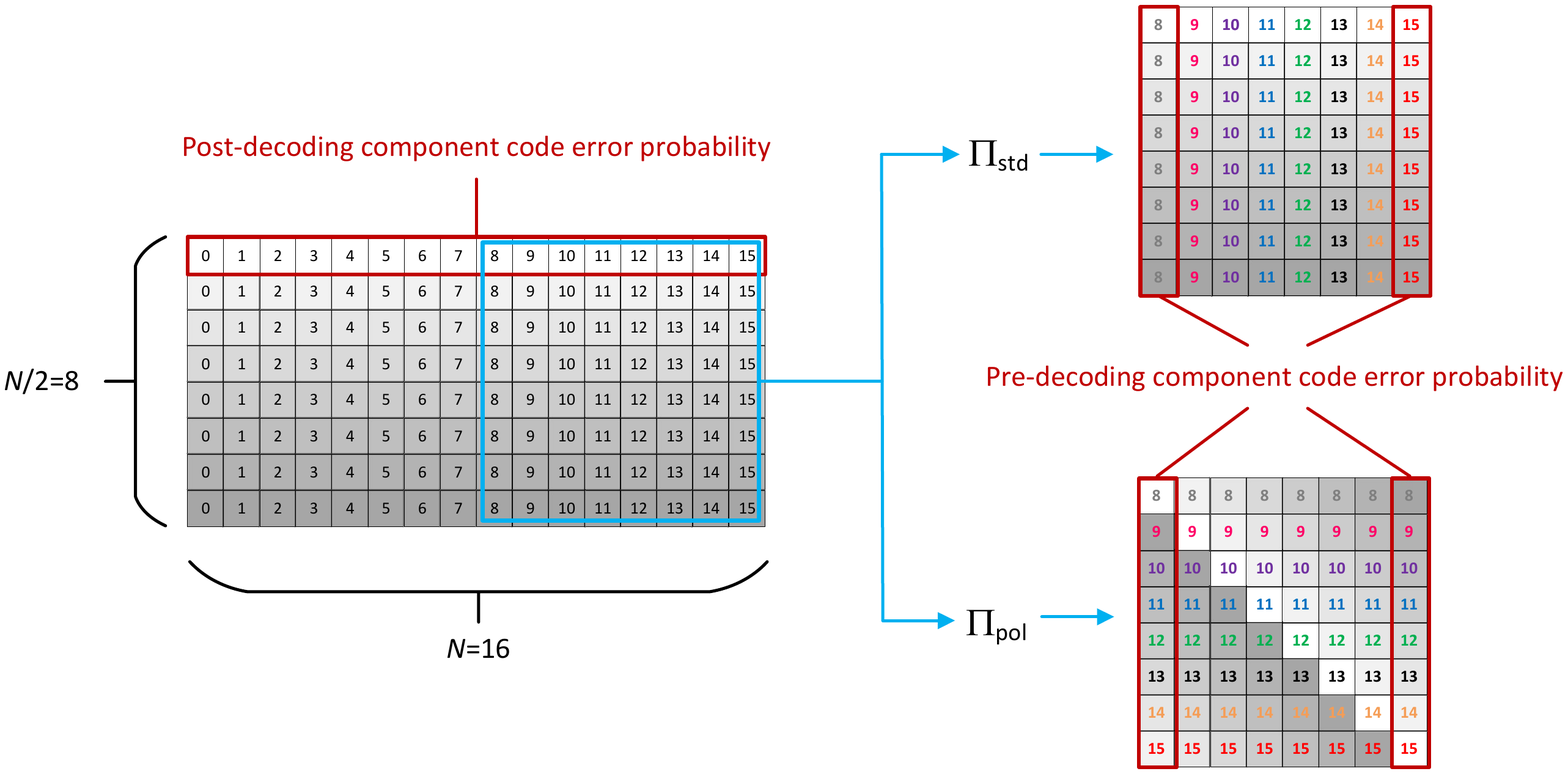}
	\caption{Standard row-column interleaver and polar-code specific interleaver in the decoder.}
	\label{fig:INT}
\end{figure*}

\section{Interleaver design} \label{sec:interleaver}

The code structure described in Section \ref{sec:NSSPC} implies a specific interleaver  between the row and column encoding phases, i.e. a clockwise-counterclockwise $90$-degree rotation that we identify as $\Pi_{\rm std}$.
Given position $(i,j)$ in a $N/2 \times N/2$ matrix $\boldsymbol{X}'$ encoded by rows, with $0\le i,j < N/2$, the row-to-column interleaver part of $\Pi_{\rm std}$ ($\Pi_{RtoC}$) is equivalent to a 90-degree clockwise rotation, given by
\begin{equation}
x_{\rm col}(j,N/2-1-i)=x_{\rm row}(i,j)~,
\end{equation}
while in the opposite direction, the column-to-row $\Pi_{CtoR}$ 90-degree counterclockwise rotation is expressed as
\begin{equation}
x_{\rm row}(N/2-1-j,i)=x_{\rm col}(i,j)~.
\end{equation}
However, any 1-to-1 interleaver can be used in the encoding process. 

Staircase constructions with algebraic component codes rely on the fact that all bits in the component codewords have the same probability of error. 
The standard row-column interleaver endemic to staircase constructions does not take in account that serial decoding algorithms, like those based on SC, return estimated vectors with unevenly distributed probability of errors.
Since the estimation of bit $i$ relies on the correct estimation of previous bits $j<i$, its probability of error is approximately
\begin{equation}
P_e^i \lessapprox P_{1e}^{i}  + \frac{1}{2} \sum_{j=0}^{i-1}P_{1e}^j~,
\end{equation}
where $P_{1e}^{j}$ is the probability of making the first decoding mistake when estimating bit $j$, as determined during the frozen set design.
Figure \ref{fig:INT} considers the example of component codewords of length $N=16$, where bits are numbered in ascending order of error probability. 
After a decoding phase (row decoding), the row-column interleaver $\Pi_{\rm std}$ distributes the resulting error probabilities so that for the following decoding phase (column decoding) each component codeword is composed of bits with the same $P_e$. 
This leads to large variations in the probability of errors among different codewords, which can negatively impact the overall error-correction capability of the staircase code.

To avoid such drawback, we propose an alternative interleaver that caters specifically to the particularities of polar codes, identified by $\Pi_{\rm pol}$ in Figure \ref{fig:INT}. 
As with $\Pi_{\rm std}$, each bit of the interleaved half-codeword is taken from a different non-interleaved half-codeword. 
However, $\Pi_{\rm pol}$ ensures that each bit has a different $P_e$, and that the cumulative $P_e$ is constant in each interleaved codeword. 
Within an interleaved half-codeword, different permutations of the interleaved bits are possible to guarantee a constant cumulative $P_e$. 
Beside $\Pi_{\rm std}$, three $\Pi_{\rm pol}$ interleavers have been evaluated, termed  $\Pi_{\rm pol}^{\rm trs}$, $\Pi_{\rm pol}^{\rm frz}$, and $\Pi_{\rm pol}^{\rm rnd}$.

The $\Pi_{\rm pol}^{\rm trs}$ interleaver performs a transposition of the interleaved matrix.
This interleaver is the one shown in Figure \ref{fig:INT}, and its function can be expressed as 
\begin{equation}
x_{\rm \Pi}(i,j)=x\big((i-j)\mod(N/2),j\big)~.
\end{equation}

\begin{figure}[t!]
	\centering
	\includegraphics[width=\linewidth]{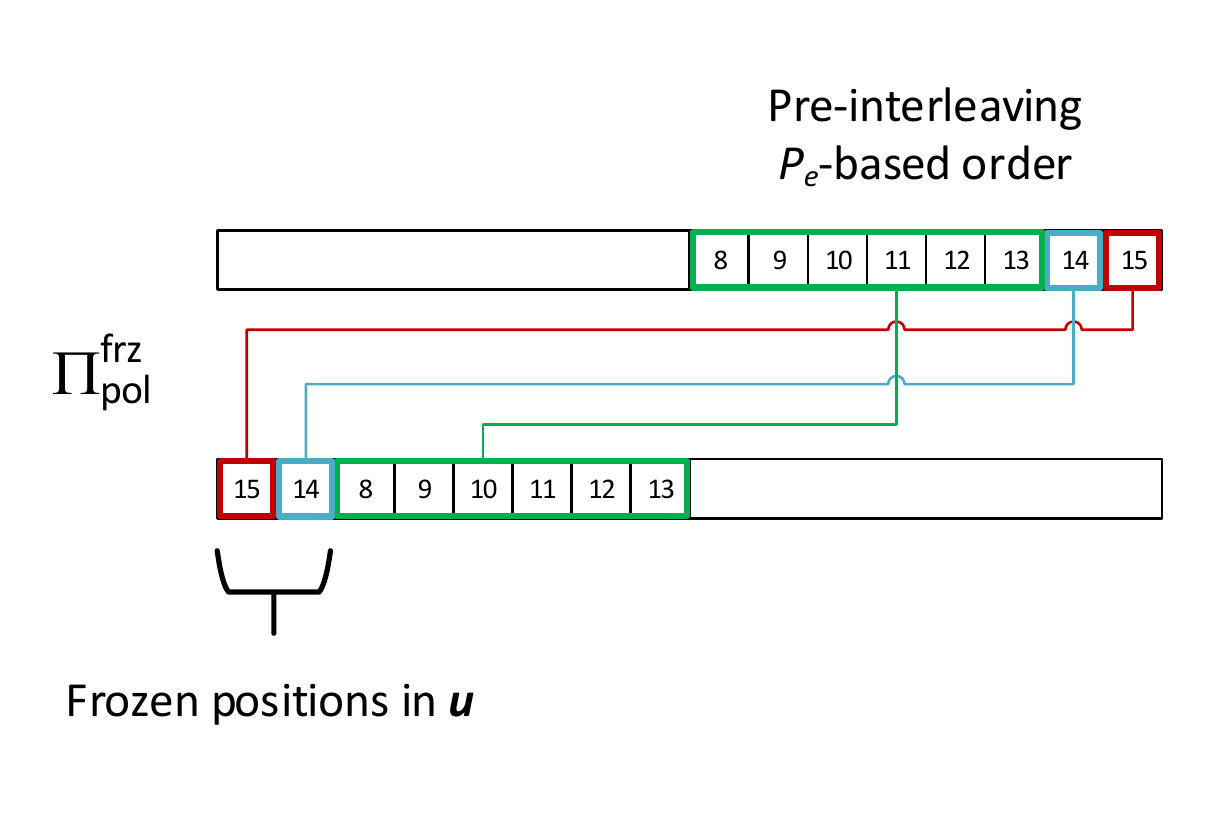}
	\vspace{-30pt}
	\caption{$\Pi_{\rm pol}^{\rm frz}$ interleaver for $N=16$, $K=14$.}
	\label{fig:frz}
\end{figure}

The $\Pi_{\rm pol}^{\rm frz}$ interleaver exploits the fact that, according to \eqref{eq:G} --\eqref{eq:F}, and by extension to all the SC-based decoders detailed in Section \ref{subsec:dec}, each channel LLR $\lambda^{(n)}_i$, corresponding to code bit $x_i$, directly contributes to the estimation of some of the source bits $u_j$ where $j\le i$. 
Assuming that all $u_j$ with $j<i$ have been correctly estimated, $u_i$ will be estimated incorrectly if $\lambda^{(n)}_i$ has the wrong sign and a large enough magnitude. 
The probability of this occurrence increases with the post-decoding $P_e$ depicted in Figure \ref{fig:INT}.
However, it is not possible to incorrectly estimate a frozen bit; LLRs with high $P_e$ can thus be neutralized by interleaving them onto frozen positions.
After $N-K$ positions have been filled, the remaining positions are assigned to the $K$ bits with the lowest $P_e$, in ascending order, again to minimize the possibility of incorrect estimation.
An example is shown in Figure \ref{fig:frz}.

Finally, $\Pi_{\rm pol}^{\rm rnd}$ performs a random permutation of the bits assigned to the interleaved half-codeword.

Simulation results considering the four described interleavers under different decoding conditions are provided in Section \ref{sec:simres}.

\section{Simulations and results}\label{sec:simres}

To evaluate the impact of various design parameters on the error-correction performance of the proposed FEC scheme, a wide-ranging simulation campaign has been performed over the AWGN channel, considering BPSK modulation. 
We observed the effect of component code block length and rate, staircase decoding window size, interleaver, and decoding algorithm parameters, like list size and internal iterations.

\begin{figure}[t!]
  \centering
   \scalebox{1}{\input{Interleaver_BER.tikz}}
   \ref{IL}
  \\
  \vspace{2pt}
  \caption{BER curves for $N=256$, $R=5/6$, $W=10$, no CRC, with soft-SCL $L=8$, SCAN $T=1$, SCANL $L=8$ and $T=1$, and different interleavers.}
  \label{fig:ILBER}
\end{figure}
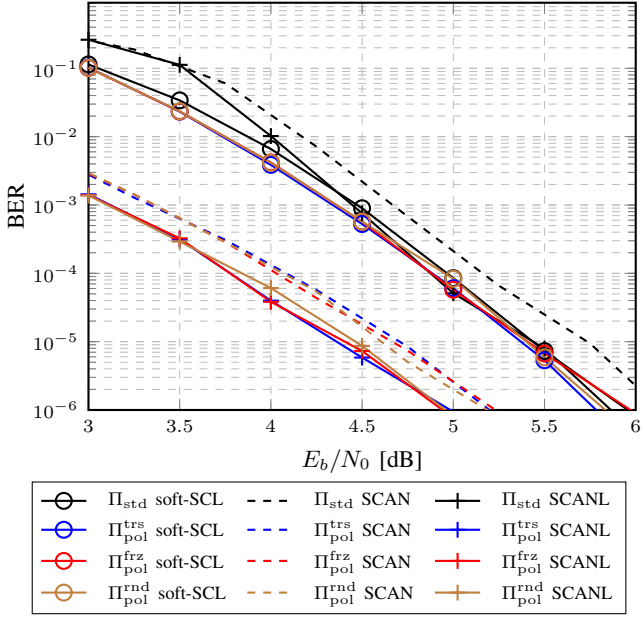


The error-correction performance of the soft-SCL decoding algorithm tends to improve as the list size $L$ increases, under the condition that the $N/L$ ratio is not too large.
Soft-SCL relies on the differences among the $L$ decoding candidates to compute the soft output; as noted in \cite{PPCjournal}, a small $L$ leads to many soft values being the same, thus sensitivity with which different reliabilities can be expressed, and thus degrading performance. 
The SCAN decoder is an iterative decoder and, as such, its performance can be expected to improve with the number of internal iterations $T$.
While this is true, the observed BER gain decreases as $T$ keeps increasing, up to the point that each additional iteration has a negligible effect. 
This effect can be traced back to the fact that if $T=1$ the soft information at the output of SCAN is guaranteed to be extrinsic, while successive degrees of approximations are applied for $T>1$. 
The same observation has been made for SCANL independently from the chosen size of the permutation space $L$, whose impact on the error-correction performance is different.
The decoding output of SCAN-based permuted graph decoding like that of SCANL does not guarantee a high degree of candidate variety; unlike true list decoders, like SCL, there is no guarantee that the $L$ results will be different.
Consequently, for very small values of $L>1$, there is no perceivable gain with respect to SCAN with the same $T$. 
While it can be expected that the improvement brought by increasing $L$ reduces its gain at high values of $L$, much like with $T$, this effect is observed at very high $L$, where the main concern becomes the implementation complexity of such decoders. 
The performance of the component decoder within the staircase framework has to be analyzed also in terms of quality of the soft output. 
SCAN and SCANL are naturally able to return a Gaussian distribution of soft values; this feature helps a smooth inter-codeword message passing.
In soft-SCL instead, many soft outputs are bound to be equal, since any bit without competitor is assigned the same value (\ref{eq:soft_nocandidate}); the resulting distribution, at least for small list sizes, is not a good approximation of a Gaussian.


The impact of the different interleavers detailed in Section \ref{sec:interleaver} has been evaluated, and the BER gain brought by each of them is portrayed in Figure \ref{fig:ILBER}. 
The simulated code considers a decoding window of $W=10$ frames, and component codes with $N=256$, $R=5/6$. 
It has been decoded with the three decoding algorithms detailed in Section \ref{subsec:compdec}; soft-SCL with $L=8$, SCAN with $T=1$, and SCANL with $L=8$ and $T=1$. 
It can be seen that the interleaver has different impacts on the BER depending on the decoding algorithm.
The different permutations have similar impact on the BER of SCAN and SCANL, improving the performance of approximately $1.8$dB BER=$10^{-6}$.
On the other hand, the error-correction performance of soft-SCL with $\Pi_{\rm std}$ and $\Pi_{\rm pol}^{\rm frz}$ is very similar, while an improvement of $0.2$dB is observed at BER=$10^{-6}$ with $\Pi_{\rm pol}^{\rm trs}$. 
This is due to the fact that (\ref{eq:soft})-(\ref{eq:soft_nocandidate}) result in a distribution of LLRs that is very different from that inferred by SCAN and SCANL, and for which $\Pi_{\rm std}$ is less suboptimal. 

For the remainder of this Section, we present results obtained with $L=8$ for both soft-SCL and SCANL, and $T=4$ for both SCAN and SCANL. 
Moreover, all simulations have been run with the $\Pi_{\rm pol}^{\rm frz}$ interleaver.
All codes have been decoded with the $\alpha_E$ (and $\alpha_B$ in case of soft-SCL) that minimizes the $E_b/N_0$ at BER=$10^{-7}$, obtained via simulation.

\begin{figure}[t!]
  \centering
   \scalebox{1}{\input{BER_CRC.tikz}}
   \ref{BER_CRC}
  \\
  \vspace{2pt}
  \caption{BER and \% of decoded codewords with $N=128$, $R=7/8$, $W=10$, no CRC, for different CRC and DR choices.}
  \label{fig:BER_CRC}
\end{figure}
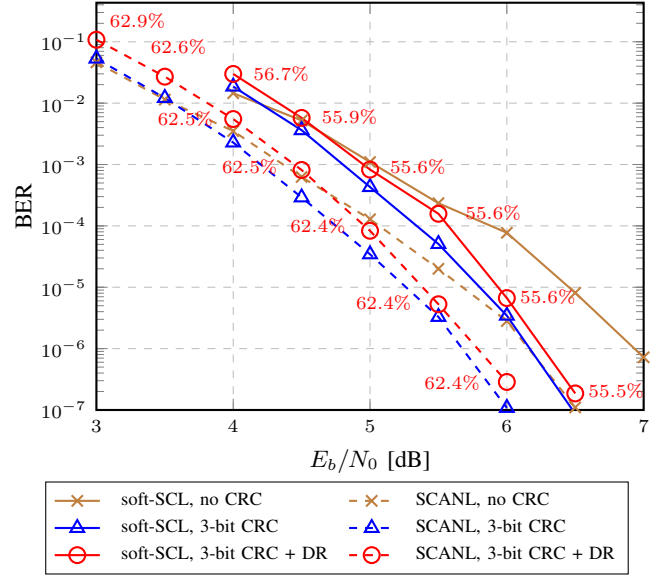


The concatenation of a CRC with the polar code helps the soft-SCL and SCANL component decoders to identify the correct candidate, thus improving the error-correction performance. 
Figure \ref{fig:BER_CRC} plots the BER in case of $N=128$, $R=7/8$, $W=10$, together with the percentage of decoded polar codes in case the decoding reduction technique described in Section \ref{subsec:decred} is activated (curves labeled as DR). 
A 3-bit CRC is used in the appropriate cases. 
It can be seen that CRC concatenation allows soft-SCL to improve its decoding performance of $0.75$dB at BER=$10^{-6}$, while a gain of $0.5$dB is observed in case of SCANL. 
This is consistent with the observations in \cite{SCANL}. 
Decoding reduction incurs a BER degradation of 0.1dB-0.15dB with respect to the standard soft-SCL and SCAN with concatenated CRC. 
However, only a fraction of polar codes are decoded; the percentage in case of soft-SCL is between 57\% and 55\%, while it stays around between 63\% and 62\% in case of SCANL. 
While SCAN can benefit from the decoding reduction, no advantage is brought by the CRC concatenation, that incurs a slight rate loss.
Similar effects have been observed with different $W$ and component decoder parameters. 
These results imply that the standard sliding-window schedule performs a certain number of unnecessary component decodings, since almost the same BER can be obtained with only slightly more than half the polar decoding attempts. 
Nevertheless, the identification of which component codes need to be decoded is a delicate operation: as shown later in this Section, simply reducing the decoding window size can in fact have a disastrous impact on the error-correction performance.


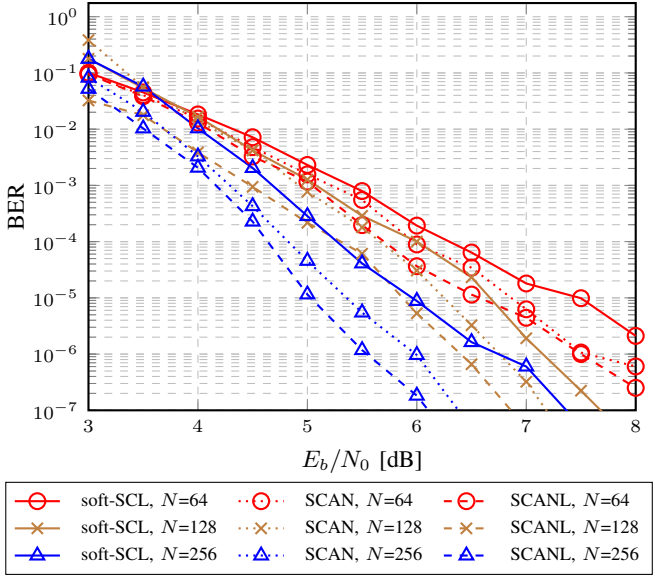
\begin{figure}[t!]
  \centering
   \scalebox{1}{\input{BER_N.tikz}}
   \ref{BER_N}
  \\
  \vspace{2pt}
  \caption{BER with different $N$ for $R=7/8$, $W=5$, no CRC.}
  \label{fig:BER_N}
\end{figure}

The size of the component decoder is a very important factor, as it affects the local error correction capability, and the number of decoding and message passing hops between two bits in the decoding window. 
Figure \ref{fig:BER_N} shows the performance of the three considered decoding algorithms at the variation of the component code length $N$, for $R=7/8$ and $W=5$.
In all observed cases, the BER curves improve with larger $N$, and as the $N$ increases SCAN and SCANL bring larger gain on soft-SCL. 
This is in accordance with the observations made earlier on the soft output of soft-SCL; by rising $N$ and keeping $L$ unchanged, a high percentage of soft messages might end up being the same. 
At $N=64$, the gain brought by SCAN and SCANL is generally smaller; moreover, SCANL improves the performance of SCAN only marginally.

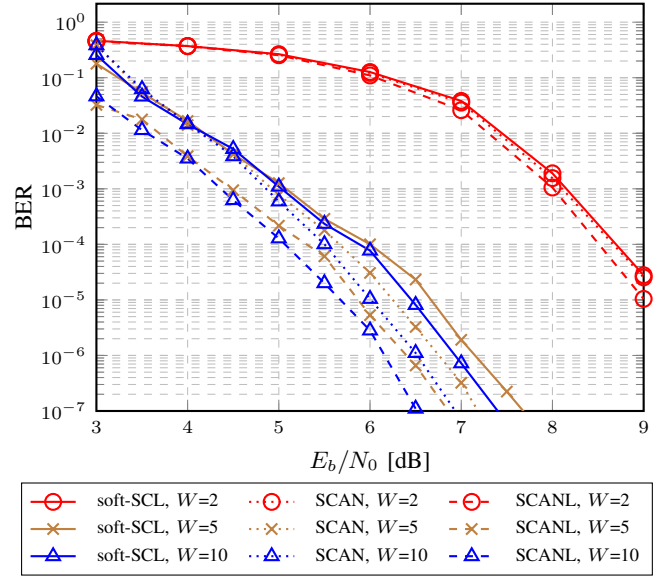
\begin{figure}[t!]
  \centering
   \scalebox{1}{\input{BER_W.tikz}}
   \ref{BER_W}
  \\
  \vspace{2pt}
  \caption{BER with different $W$ for $N=128$, $R=7/8$, no CRC.}
  \label{fig:BER_W}
\end{figure}

The dimension of the decoding window $W$ has a large impact on the error-correction performance, as it determines how many frames can directly exchange information, as well as greatly impacting decoding latency and implementation complexity.
In general, a larger $W$ tends to improve the BER, and increase latency and cost metrics.
Figure \ref{fig:BER_W} portrays the BER as the window size $W$ changes. 
When decoded with soft-SCL, the BER improves of more than $2.5$dB when increasing $W$ from 2 to 5; the larger decoding window allows to correct a higher number of errors, and the waterfall region of the curves starts at lower $E_b/N_0$. 
However, within these decoding conditions, increasing $W$ from $5$ to $10$ does not improve the error correction performance, due to the very local nature of $\alpha_E$ and $\alpha_B$ (\ref{eq:soft})-(\ref{eq:soft_nocandidate}), that need to be varied across the decoding window as $W$ increases. 
In case of SCAN and SCANL this effect is more mitigated, as only $\alpha_E$ is used; consequently, a more substantial gain ($\approx0.4$dB) is observed when increasing $W$ from 5 to 10.
Moreover, both SCAN and SCANL at $W=5$ result in more than $3$dB gain compared to $W=2$ with the same algorithm.
With respect to soft-SCL, SCAN and SCANL yield $0.5$dB to $1$dB at $W=5$ and $W=10$, but very small improvements for $W=2$, where the size of the decoding window is too small to exploit the improved soft information.

In \cite{Staircase_polar2}, systematic polar codes are used in a staircase construction, and a soft decoding algorithm similar to soft-SCL is proposed, that is shown to perform better than some versions of belief propagation and SCAN decoding. 
The staircase code is built with partial codeword superposition, using $M<N/2$. 
Simulation results with $W=5$, $N=256$, $M=80$, $R=7/8$, $L=32$ are shown, achieving a BER=$6\cdot10^{-6}$ at $E_b/N_0=5.5$. 
Similar decoding conditions are found in Figure \ref{fig:BER_N}, however with $M=128$ and $L=8$.
We can see that while our version of soft-SCL is not able to make up for the difference in $L$, SCAN is able to match the performance of \cite{Staircase_polar2}, and SCANL can improve it of approximately $0.3$dB. 
The performance of \cite{Staircase_polar2} are matched by soft-SCL with $L=8$ by including a 4-bit CRC in each component code, while under the same conditions SCANL yields a $0.9$dB gain.

\section{Conclusion}\label{sec:conc}
In this work, we have presented a staircase code construction with non-systematic polar codes, detailing encoding and decoding methods. 
Component codes are decoded with one of three decoding algorithms: soft-SCL, SCAN, and SCANL. 
We proposed to concatenate the component codes with an outer CRC code; this feature brings $0.5$dB-$0.75$dB gain when decoded with soft-SCL or SCANL at BER=$10^{-6}$. 
Moreover, we propose to use successful CRC passes to reduce the decoding load; simulation results show up to $44.5\%$ fewer component decoding attempts, at the cost of $0.1$dB BER degradation at BER=$10^{-6}$. 
To cater to the specific nature of polar codes, we designed ad-hoc interleavers that consider their uneven distribution of error probabilities, and we observed up to $0.9$dB gain with SCAN and SCANL decoding with respect to the standard staircase interleaver. 
We finally presented simulation results investigating the error-correction performance of the proposed scheme at the variation of multiple code and decoding parameters. 
Under the same parameters, we can match the best performance of staircase construction with systematic polar component codes in literature with a fraction of the decoding complexity, or gain approximately $0.9$dB with the same order of complexity.

\end{document}

%% file: G_8.tex
\begin{tikzpicture}

\draw  (-1.5,0.5) -- (12.5,0.5) ;
\draw  (-1.5,-1) -- (12.5,-1);
\draw  (0.5,0.5) ellipse (.5 and .5);
\draw  (0.5,1) -- (0.5,-1);
\node  at (-3,0.5) {\huge $u_0$};
\node  at (-3,-1) {\huge $u_1$};
\node  at (-3,-2.5) {\huge $u_2$};
\node at (-3,-4) {\huge $u_3$};
\node  at (-3,-5.5) {\huge $u_4$};
\node at (-3,-7) {\huge $u_5$};
\node at (-3,-8.5) {\huge $u_6$};
\node at (-3,-10) {\huge $u_7$};
\node at (14,0.5) {\huge $x_0$};
\node at (14,-1) {\huge $x_1$};
\node at (14,-2.5) {\huge $x_2$};
\node at (14,-4) {\huge $x_3$};
\node at (14,-5.5) {\huge $x_4$};
\node at (14,-7) {\huge $x_5$};
\node at (14,-8.5) {\huge $x_6$};
\node at (14,-10) {\huge $x_7$};


\draw (-1.5,-4) -- (12.5,-4);
\draw (-1.5,-2.5) -- (12.5,-2.5);
\draw (-1.5,-5.5) -- (12.5,-5.5);
\draw (-1.5,-7) -- (12.5,-7);
\draw (-1.5,-8.5) -- (12.5,-8.5);
\draw (-1.5,-10) -- (12.5,-10);
\draw (0.5,-4) -- (0.5,-2);
\draw (0.5,-7) -- (0.5,-5);
\draw (0.5,-10) -- (0.5,-8);
\draw  (0.5,-2.5) ellipse (0.5 and 0.5);
\draw  (0.5,-5.5) ellipse (0.5 and 0.5);
\draw  (0.5,-8.5) ellipse (0.5 and 0.5);
\draw (3.5,-4) -- (3.5,-0.5);
\draw (4.5,-2.5) -- (4.5,1) node (v1) {};
\draw  (3.5,-1) ellipse (0.5 and 0.5);
\draw  (4.5,0.5) ellipse (0.5 and 0.5);
\draw (3.5,-6.5) -- (3.5,-10);
\draw (4.5,-5) -- (4.5,-8.5);
\draw  (4.5,-5.5) ellipse (0.5 and 0.5);
\draw  (3.5,-7) ellipse (0.5 and 0.5);
\draw (7.5,-10) -- (7.5,-3.5);
\draw (8.5,-8.5) -- (8.5,-2);
\draw (9.5,-7) -- (9.5,-0.5);
\draw (10.5,-5.5) -- (10.5,1);
\draw [fill=white] (0.5,-10) ellipse (.1 and .1);
\draw [fill=white] (0.5,-1) ellipse (.1 and .1);
\draw [fill=white] (0.5,-4) ellipse (.1 and .1);
\draw [fill=white] (0.5,-7) ellipse (.1 and .1);
\draw [fill=white] (3.5,-10) ellipse (.1 and .1);
\draw [fill=white] (4.5,-2.5) ellipse (.1 and .1);
\draw [fill=white] (3.5,-4) ellipse (.1 and .1);
\draw [fill=white] (4.5,-8.5) ellipse (.1 and .1);
\draw [fill=white] (0.5,-7) ellipse (.1 and .1);
\draw [fill=white] (7.5,-10) ellipse (.1 and .1);
\draw [fill=white] (10.5,-5.5) ellipse (.1 and .1);
\draw [fill=white] (9.5,-7) ellipse (.1 and .1);
\draw [fill=white] (8.5,-8.5) ellipse (.1 and .1);
\draw  (7.5,-4) ellipse (0.5 and 0.5);
\draw  (8.5,-2.5) ellipse (0.5 and 0.5);
\draw  (9.5,-1) ellipse (0.5 and 0.5);
\draw  (10.5,0.5) ellipse (0.5 and 0.5);

\draw  (-0.5,1.5) rectangle (1.5,-10.5);
\draw  (2.5,1.5) rectangle (5.5,-10.5);
\draw  (6.5,-10.5) rectangle (11.5,1.5);
\end{tikzpicture}

%% file: factor-graph-2-soft.tex
\begin{tikzpicture}
\newcommand{\polarcodetikz}[7]{
\draw (#1,#2) -- (#5-#6, #2);
\draw (#5,#2) circle(#6cm); 
\draw (#5-#6,#2) -- (#5+#6,#2); 
\draw (#5,#2-#6) -- (#5,#2+#6); 
\draw (#3,#4) -- (#5-#6/3,#4);
\draw [fill=black] (#5-#6/3,#4-#6/3) rectangle (#5+#6/3,#4+#6/3);
\draw (#5,#4+#6/3) -- (#5,#2-#6);
\draw(#5+#6,#2) -- (#5+#6+#7,#2);
\draw(#5+#6/3,#4) -- (#5+#6+#7,#4);
}
\def\Rad{0.15}
\def\gap{1}
\def\out{1.25}
\draw  (0,0) node [left] {\textcolor{red}{$\Lm_a$}, \textcolor{blue}{$\B_a$}};
\draw  (0,-\gap) node [left] {\textcolor{red}{$\Lm_b$}, \textcolor{blue}{$\B_b$}};
\draw  (2*\out+0.1,0) node [right] {\textcolor{red}{$\Lm_c$}, \textcolor{blue}{$\B_c$}};
\draw  (2*\out+0.1,-\gap) node [right] {\textcolor{red}{$\Lm_d$}, \textcolor{blue}{$\B_d$}};
\polarcodetikz{0}{0}{0}{0-\gap}{\out}{\Rad}{\out}
\end{tikzpicture}

%% file: Interleaver_BER.tikz
\begin{tikzpicture}
  \pgfplotsset{
    label style = {font=\fontsize{9pt}{7.2}\selectfont},
    tick label style = {font=\fontsize{7pt}{7.2}\selectfont}
  }

\begin{axis}[
	scale = 1,
    ymode=log,
    xlabel={$E_b/N_0$ [\text{dB}]}, xlabel style={yshift=0.4em},
    ylabel={BER}, ylabel style={yshift=-0.75em},
    grid=both,
    ymajorgrids=true,
    xmajorgrids=true,
    grid style=dashed,
    mark options=solid,
    width=1\columnwidth, height=7cm,
    thick,
        xmin=3,
        xmax=6,
        ymin=1e-6,
    mark size=3,
    legend style={
      anchor={center},
      cells={anchor=west},
      mark options=solid,
      column sep= 2mm,
      font=\fontsize{7pt}{7.2}\selectfont,
    },
    legend to name=IL,
    legend columns=3,
]

\addplot[
    color=black,
    mark=o,
    thick,
    mark size=3,
]
table {
3 0.114139
3.5 0.0339745
4 0.00659681
4.5 0.000896116
5 8.50302e-05
5.5 7.29264e-06
6 8.27959e-07
};
\addlegendentry{$\Pi_{\rm std}$ soft-SCL}

\addplot[
    color=black,
    thick,
    dashed,
    mark size=3,
]
table {
1.75 0.415717
2.25 0.390315
2.75 0.341915
3.25 0.189513
3.75 0.0603753
4.25 0.0071417
4.75 0.000679368
5.25 6.59441e-05
5.75 9.40621e-06
6.25 5.42223e-07
};
\addlegendentry{$\Pi_{\rm std}$ SCAN}

\addplot[
    color=black,
    mark=+,
    thick,
    mark size=3,
]
table {
3 0.26212
3.5 0.11248
4 0.010209
4.5 0.00075591
5 5.10731e-05
5.5 7.9238e-06
6.0 4.62434e-07
};
\addlegendentry{$\Pi_{\rm std}$ SCANL}

\addplot[
    color=blue,
    mark=o,
    thick,
    mark size=3,
]
table {
3 0.10259
3.5 0.0232671
4 0.00386648
4.5 0.000526924
5 6.0544e-005
5.5 5.33622e-006
6 2.77113e-007
};
\addlegendentry{$\Pi_{\rm pol}^{\rm trs}$ soft-SCL}

\addplot[
    color=blue,
    thick,
    dashed,    
    mark size=3,
]
table {
1.75 0.112581
2.25 0.0270576
2.75 0.00604985
3.25 0.0012812
3.75 0.000308939
4.25 5.70243e-05
4.75 8.44395e-06
5.25 7.80896e-07
};
\addlegendentry{$\Pi_{\rm pol}^{\rm trs}$ SCAN}

\addplot[
    color=blue,
   mark=+,
    thick,
    mark size=3,
]
table {
2.5 0.0058395
3 0.00143572
3.5 0.000313947
4 3.98653e-05
4.5 5.85036e-06
5 9.42008e-07
};
\addlegendentry{$\Pi_{\rm pol}^{\rm trs}$ SCANL}

\addplot[
    color=red,
    mark=o,
    thick,
    mark size=3,
]
table {
3 0.101662
3.5 0.0234007
4 0.00419324
4.5 0.000569189
5 5.75157e-005
5.5 6.71346e-006
6 8.90219e-007
};
\addlegendentry{$\Pi_{\rm pol}^{\rm frz}$ soft-SCL}

\addplot[
    color=red,
    thick,
    dashed,
    mark size=3,
]
table {
1.75 0.11223
2.25 0.0268434
2.75 0.00584961
3.25 0.00142394
3.75 0.0002877
4.25 4.34483e-05
4.75 7.37889e-06
5.25 9.23401e-07
};
\addlegendentry{$\Pi_{\rm pol}^{\rm frz}$ SCAN}

\addplot[
    color=red,
    mark=+,
    thick,
    mark size=3,
]
table {
2.5 0.00594561
3 0.00139342
3.5 0.000325633
4 3.83081e-05
4.5 7.30378e-06
5 7.66031e-07
};
\addlegendentry{$\Pi_{\rm pol}^{\rm frz}$ SCANL}

\addplot[
    color=brown,
    mark=o,
    thick,
    mark size=3,
]
table {
3 0.101541
3.5 0.023566
4 0.00413741
4.5 0.000571034
5 8.32712e-005
5.5 5.90666e-006
6 4.11874e-007
};
\addlegendentry{$\Pi_{\rm pol}^{\rm rnd}$ soft-SCL}

\addplot[
    color=brown,
    thick,
    dashed,
    mark size=3,
]
table {
1.75 0.113297
2.25 0.0284148
2.75 0.00608287
3.25 0.00142959
3.75 0.000286123
4.25 5.41145e-05
4.75 5.03573e-06
5.25 7.89641e-07
};
\addlegendentry{$\Pi_{\rm pol}^{\rm rnd}$ SCAN}

\addplot[
    color=brown,
    mark=+,
    thick,
    mark size=3,
]
table {
2.5 0.00601257
3 0.0013809
3.5 0.000292059
4 6.08597e-05
4.5 8.66842e-06
5 8.14366e-07
};
\addlegendentry{$\Pi_{\rm pol}^{\rm rnd}$ SCANL}

\end{axis}
\end{tikzpicture}

%% file: BER_CRC.tikz
\begin{tikzpicture}
  \pgfplotsset{
    label style = {font=\fontsize{9pt}{7.2}\selectfont},
    tick label style = {font=\fontsize{7pt}{7.2}\selectfont}
  }

\begin{axis}[
	scale = 1,
    ymode=log,
    xlabel={$E_b/N_0$ [\text{dB}]}, xlabel style={yshift=0.4em},
    ylabel={BER}, ylabel style={yshift=-0.75em},
    ymajorgrids=true,
    xmajorgrids=true,
    grid style=dashed,
    mark options=solid,
    width=1\columnwidth, height=7cm,
    thick,
       xmin=3,
       xmax=7,
        ymin=1e-7,
    mark size=3,
    legend style={
      anchor={center},
      cells={anchor=west},
      mark options=solid,
      column sep= 2mm,
      font=\fontsize{7pt}{7.2}\selectfont,
    },
    legend to name=BER_CRC,
    legend columns=2,
]

\addplot[
    color=brown,
    mark=x,
    thick,
    mark size=3,
]
table {
4 0.0145333
4.5 0.00521732
5 0.00109665
5.5 0.00023409
6 7.72149e-05
6.5 8.11682e-06
7 7.21955e-07
7.5 6.10802e-08
};
\addlegendentry{soft-SCL, no CRC}

\addplot[
    color=brown,
    mark=x,
    dashed,
    thick,
    mark size=3,
]
table {
3   0.0462215
3.5 0.01141565
4   0.0035097
4.5 0.00062602
5   1.28110e-04
5.5 1.999238e-05
6   2.81333e-06
6.5 1.100222e-07
};
\addlegendentry{SCANL, no CRC}

\addplot[
    color=blue,
    mark=triangle,
    thick,
    mark size=3,
]
table {
4 0.018622		
4.5 0.00362245
5 0.000428991 
5.5 5.06714e-005
6 3.44192e-006
6.5 8.658521e-008
};
\addlegendentry{soft-SCL, 3-bit CRC}

\addplot[
  color=blue,
    mark=triangle,
	dashed,
    thick,
    mark size=3,
]
table {
3   0.0531567
3.5 0.0121376
4   0.0022833
4.5 0.00029101
5   3.44313e-05
5.5 3.28555e-06
6   1.081333e-07
};
\addlegendentry{SCANL, 3-bit CRC}

\addplot[
    color=red,
    mark=o,
    thick,
    mark size=3,
]
table {
4 0.0298947 
4.5 0.00574638	
5 0.000827434 
5.5 0.000157492
6 6.66318e-006
6.5 1.8607e-007
};
\addlegendentry{soft-SCL, 3-bit CRC + DR}
\draw [red] (135,-3.5)  node  {\scriptsize $56.7\%$};
\draw [red] (185,-5.1)  node  {\scriptsize $55.9\%$};
\draw [red] (235,-7)  node  {\scriptsize $55.6\%$};
\draw [red] (290,-8.7)  node  {\scriptsize $55.6\%$};
\draw [red] (330,-11.85)  node  {\scriptsize $55.6\%$};
\draw [red] (380,-15.4)  node  {\scriptsize $55.5\%$};

\addplot[
	color=red,
    mark=o,
	dashed,
    thick,
    mark size=3,
]
table {
3   0.107873
3.5 0.0270051
4   0.0055108
4.5 0.00081392
5   8.3288e-05
5.5 5.28555e-06
6   2.861472e-07
};
\addlegendentry{SCANL, 3-bit CRC + DR}
\draw [red] (30,-1.5)  node  {\scriptsize $62.9\%$};
\draw [red] (60,-2.5)  node  {\scriptsize $62.6\%$};
\draw [red] (65,-5.2)  node  {\scriptsize $62.5\%$};
\draw [red] (112,-7.0)  node  {\scriptsize $62.5\%$};
\draw [red] (162,-9.2)  node  {\scriptsize $62.4\%$};
\draw [red] (210,-12.1)  node  {\scriptsize $62.4\%$};
\draw [red] (260,-15.1)  node  {\scriptsize $62.4\%$};

\end{axis}
\end{tikzpicture}

%% file: BER_N.tikz
\begin{tikzpicture}
  \pgfplotsset{
    label style = {font=\fontsize{9pt}{7.2}\selectfont},
    tick label style = {font=\fontsize{7pt}{7.2}\selectfont}
  }

\begin{axis}[
	scale = 1,
    ymode=log,
    xlabel={$E_b/N_0$ [\text{dB}]}, xlabel style={yshift=0.4em},
    ylabel={BER}, ylabel style={yshift=-0.75em},
    grid=both,
    ymajorgrids=true,
    xmajorgrids=true,
    grid style=dashed,
    mark options=solid,
    width=1\columnwidth, height=7cm,
    thick,
        xmin=3,
        xmax=8,
        ymin=1e-7,
    mark size=3,
    legend style={
      anchor={center},
      cells={anchor=west},
      mark options=solid,
      column sep= 2mm,
      font=\fontsize{7pt}{7.2}\selectfont,
    },
    legend to name=BER_N,
    legend columns=3,
]

\addplot[
    color=red,
    mark=o,
    thick,
    mark size=3,
]
table {
3 0.10096 		
3.5 0.0450487 		
4 0.0182962 	
4.5 0.00717224 	
5 0.00230937		 
5.5 0.000792558 	
6 0.000192829 	
6.5 6.39715e-005 	
7 1.80176e-005 	
7.5 9.85621e-006
8 2.10483e-006
};
\addlegendentry{soft-SCL, $N$=64}

\addplot[
    color=red,
    mark=o,
    thick,
    dotted,
    mark size=3,
]
table {
3   0.09545
3.5 0.04023
4   0.0154521
4.5 0.0047922
5   0.0015711
5.5 0.00055141
6   0.0000890351
6.5 3.436602e-005	
7   6.25766e-006 	
7.5 1.066391e-006
8   6.03076e-007
};
\addlegendentry{SCAN, $N$=64}

\addplot[
    color=red,
    mark=o,
    thick,
    dashed,
    mark size=3,
]
table {
3   0.09485
3.5 0.03853
4   0.012995
4.5 0.003271
5   0.0011693
5.5 0.00019655
6   0.000036522
6.5 1.14407e-005	
7   4.419028e-006 	
7.5 1.000422e-006
8   2.510845e-007
};
\addlegendentry{SCANL, $N$=64}

\addplot[
    color=brown,
    mark=x,
    thick,
    mark size=3,
]
table {
3 0.17865
3.5 0.053722
4 0.0166311
4.5 0.0042098
5 0.0012656
5.5 0.00028778
6 0.000100351
6.5 2.32227e-05
7 1.9125e-06
7.5 2.23651e-07
8 2.38401e-08
};
\addlegendentry{soft-SCL, $N$=128}

\addplot[
    color=brown,
    mark=x,
    dotted,
    thick,
    mark size=3,
]
table {
3   0.3842
3.5 0.055821
4   0.014031
4.5 0.0042006
5   0.0007724
5.5 0.0001792
6   3.05952e-05
6.5 3.26313e-06
7   3.21955e-07
7.5 1.32897e-08
};
\addlegendentry{SCAN, $N$=128}

\addplot[
    color=brown,
    mark=x,
    dashed,
    thick,
    mark size=3,
]
table {
3   0.032376
3.5 0.0177332
4   0.0038595
4.5 0.00094405
5   2.16503e-04
5.5 6.097854e-05
6   5.312388e-06
6.5 6.596802e-07
7   4.81333e-08
};
\addlegendentry{SCANL, $N$=128}

\addplot[
    color=blue,
    mark=triangle,
    thick,
    mark size=3,
]
table {
3 0.17812
3.5 0.0572031
4 0.0103441
4.5 0.00203972
5 0.000285036
5.5 4.09842e-05
6 8.80042e-06
6.5 1.6324e-06
7 6.00895e-07
7.5 5.20263e-08
};
\addlegendentry{soft-SCL, $N$=256}

\addplot[
    color=blue,
    mark=triangle,
    dotted,
    thick,
    mark size=3,
]
table {
3   0.08232
3.5 0.02033508
4   0.0032769
4.5 0.00042699
5   4.55053e-05
5.5 5.48116e-06
6   9.62867e-07
6.5 4.85343e-08
};
\addlegendentry{SCAN, $N$=256}

\addplot[
    color=blue,
    mark=triangle,
    dashed,
    thick,
    mark size=3,
]
table {
3   0.052208
3.5 0.0103101
4   0.00206358
4.5 0.00022693
5   1.149839e-05
5.5 1.17754e-06
6   1.82064e-07
6.5 1.0605511e-8
};
\addlegendentry{SCANL, $N$=256}

%

\end{axis}
\end{tikzpicture}

%% file: BER_W.tikz
\begin{tikzpicture}
  \pgfplotsset{
    label style = {font=\fontsize{9pt}{7.2}\selectfont},
    tick label style = {font=\fontsize{7pt}{7.2}\selectfont}
  }

\begin{axis}[
	scale = 1,
    ymode=log,
    xlabel={$E_b/N_0$ [\text{dB}]}, xlabel style={yshift=0.4em},
    ylabel={BER}, ylabel style={yshift=-0.75em},
    grid=both,
    ymajorgrids=true,
    xmajorgrids=true,
    grid style=dashed,
    mark options=solid,
    width=1\columnwidth, height=7cm,
    thick,
        xmin=3,
		xmax=9,
		ymin=1e-7,
    mark size=3,
    legend style={
      anchor={center},
      cells={anchor=west},
      mark options=solid,
      column sep= 2mm,
      font=\fontsize{7pt}{7.2}\selectfont,
    },
    legend to name=BER_W,
    legend columns=3,
]

\addplot[
    color=red,
    mark=o,
    thick,
    mark size=3,
]
table {
3 0.46311
4 0.37121
5 0.265327
6 0.12664
7 0.038242
8 0.00190082
9 2.77837e-005
};
\addlegendentry{soft-SCL, $W$=2}

\addplot[
    color=red,
    mark=o,
    dotted,
    thick,
    mark size=3,
]
table {
3 0.46062
4 0.37281
5 0.258404
6 0.12199
7 0.035637
8 0.0015575
9 2.53811e-005
};
\addlegendentry{SCAN, $W$=2}

\addplot[
    color=red,
    mark=o,
    dashed,
    thick,
    mark size=3,
]
table {
3 0.45202
4 0.36836
5 0.25402
6 0.10994
7 0.0260105
8 0.00105539
9 1.04205e-005
};
\addlegendentry{SCANL, $W$=2}

\addplot[
    color=brown,
    mark=x,
    thick,
    mark size=3,
]
table {
3 0.17865
3.5 0.053722
4 0.0166311
4.5 0.0042098
5 0.0012656
5.5 0.00028778
6 0.000100351
6.5 2.32227e-05
7 1.9125e-06
7.5 2.23651e-07
8 2.38401e-08
};
\addlegendentry{soft-SCL, $W$=5}

\addplot[
    color=brown,
    mark=x,
    dotted,
    thick,
    mark size=3,
]
table {
3   0.3842
3.5 0.055821
4   0.014031
4.5 0.0042006
5   0.0007724
5.5 0.0001792
6   3.05952e-05
6.5 3.26313e-06
7   3.21955e-07
7.5 1.32897e-08
};
\addlegendentry{SCAN, $W$=5}

\addplot[
    color=brown,
    mark=x,
    dashed,
    thick,
    mark size=3,
]
table {
3   0.032376
3.5 0.0177332
4   0.0038595
4.5 0.00094405
5   2.16503e-04
5.5 6.097854e-05
6   5.312388e-06
6.5 6.596802e-07
7   4.81333e-08
};
\addlegendentry{SCANL, $W$=5}

\addplot[
    color=blue,
    mark=triangle,
    thick,
    mark size=3,
]
table {
3 0.25711
3.5 0.045821
4 0.0145333
4.5 0.00521732
5 0.00109665
5.5 0.00023409
6 7.72149e-05
6.5 8.11682e-06
7 7.21955e-07
7.5 6.10802e-08
};
\addlegendentry{soft-SCL, $W$=10}

\addplot[
    color=blue,
    mark=triangle,
    dotted,
    thick,
    mark size=3,
]
table {
3   0.3843
3.5 0.06233
4   0.014874
4.5 0.0038551
5   0.0005929
5.5 0.0001006
6   1.04327e-05
6.5 1.106802e-06
7   6.972502e-08
};
\addlegendentry{SCAN, $W$=10}

\addplot[
    color=blue,
    mark=triangle,
    dashed,
    thick,
    mark size=3,
]
table {
3   0.0462215
3.5 0.01141565
4   0.0035097
4.5 0.00062602
5   1.28110e-04
5.5 1.999238e-05
6   2.81333e-06
6.5 1.100222e-07
};
\addlegendentry{SCANL, $W$=10}

\end{axis}
\end{tikzpicture}

%% file: Staircase_journal.bbl
\begin{thebibliography}{10}

\bibitem{Staircase}
B.~P. {Smith}, A.~{Farhood}, A.~{Hunt}, F.~R. {Kschischang}, and J.~{Lodge},
\newblock ``Staircase codes: {FEC for 100 Gb/s OTN},''
\newblock {\em Journal of Lightwave Technology}, vol. 30, no. 1, pp. 110--117,
  Jan 2012.

\bibitem{ITUG709.2}
International~Telecommunication Union,
\newblock ``{OTU4} long-reach interface,''
\newblock {\em ITU-T Rec. G.709.2/Y.1331.2}, 2018.

\bibitem{OIF400ZR}
Optical~Internetworking Forum,
\newblock ``Implementation agreement {400ZR},''
\newblock {\em OIF-400ZR-01.0}, 2020.

\bibitem{StairCaseWireless}
P.~{Kukieattikool} and N.~{Goertz},
\newblock ``Staircase codes for high-rate wireless transmission on burst-error
  channels,''
\newblock {\em IEEE Wireless Communications Letters}, vol. 5, no. 2, pp.
  128--131, Apr. 2016.

\bibitem{Zipper}
A.~Y. {Sukmadji}, U.~{Martínez-Peñas}, and F.~R. {Kschischang},
\newblock ``Zipper codes: Spatially-coupled product-like codes with iterative
  algebraic decoding,''
\newblock in {\em Canadian Workshop on Information Theory (CWIT)}, Hamilton,
  ON, Canada, June 2019.

\bibitem{Gallager}
R.~{Gallager},
\newblock ``Low-density parity-check codes,''
\newblock {\em IRE Transactions on Information Theory}, vol. 8, no. 1, pp.
  21--28, January 1962.

\bibitem{Urbanke/Richardson}
T.~J. {Richardson}, M.~A. {Shokrollahi}, and R.~L. {Urbanke},
\newblock ``Design of capacity-approaching irregular low-density parity-check
  codes,''
\newblock {\em IEEE Transactions on Information Theory}, vol. 47, no. 2, pp.
  619--637, Feb. 2001.

\bibitem{ArikanFirst}
E.~Arikan,
\newblock ``Channel polarization: A method for constructing capacity-achieving
  codes for symmetric binary-input memoryless channels,''
\newblock {\em IEEE Transactions on Information Theory}, vol. 55, no. 7, pp.
  3051--3073, July 2009.

\bibitem{3GPP_R15}
$3^{\text{rd}}$ Generation Partnership Project~({3GPP}),
\newblock ``Multiplexing and channel coding,''
\newblock {\em 3GPP 38.212 V.15.3.0}, 2018.

\bibitem{FastDec}
G.~{Sarkis}, P.~{Giard}, A.~{Vardy}, C.~{Thibeault}, and W.~J. {Gross},
\newblock ``Fast polar decoders: Algorithm and implementation,''
\newblock {\em IEEE Journal on Selected Areas in Communications}, vol. 32, no.
  5, pp. 946--957, 2014.

\bibitem{SCPOLARConcat}
Tayyab Mehmood, {Metodi Plamenov} Yankov, Anders Fisker, Kim Gormsen, and
  S{\o}ren Forchhammer,
\newblock ``Rate-adaptive concatenated polar-staircase codes for data center
  interconnects,''
\newblock in {\em Optical Fiber Communication Conference}, Mar. 2020.

\bibitem{par_conc_sys_pol}
D.~Wu, A.~Liu, Y.~Zhang, and Q.~Zhang,
\newblock ``Parallel concatenated systematic polar codes,''
\newblock {\em Electronics Letters}, vol. 52, no. 1, pp. 43--45, 2015.

\bibitem{KoikeAkinoIrregularPT}
T.~Koike-Akino, C.~Cao, Y.~Wang, K.~Kojima, D.~S. Millar, and K.~Parsons,
\newblock ``Irregular polar turbo product coding for high-throughput optical
  interface,''
\newblock in {\em Optical Fiber Communication Conference (OFC)}, San Diego, CA,
  USA, March 2018.

\bibitem{PPC}
V.~{Bioglio}, C.~{Condo}, and I.~{Land},
\newblock ``Construction and decoding of product codes with non-systematic
  polar codes,''
\newblock in {\em IEEE Wireless Communications and Networking Conference
  (WCNC)}, Marrakech, Morocco, Apr. 2019.

\bibitem{PPCjournal}
C.~{Condo}, V.~{Bioglio}, H.~{Hafermann}, and I.~{Land},
\newblock ``Practical product code construction of polar codes,''
\newblock {\em IEEE Transactions on Signal Processing}, vol. 68, pp.
  2004--2014, 2020.

\bibitem{Staircase_polar}
B.~{Feng}, J.~{Jiao}, L.~{Zhou}, S.~{Wu}, B.~{Cao}, and Q.~{Zhang},
\newblock ``A novel high-rate polar-staircase coding scheme,''
\newblock in {\em 2018 IEEE 88th Vehicular Technology Conference (VTC-Fall)},
  Chicago, IL, USA, Aug. 2018.

\bibitem{Staircase_polar2}
L.~{Zhou}, B.~{Feng}, J.~{Jiao}, K.~{Liang}, S.~{Wu}, and Q.~{Zhang},
\newblock ``Performance analysis of soft decoding algorithms for
  polar-staircase coding scheme,''
\newblock in {\em International Conference on Wireless Communications and
  Signal Processing (WCSP)}, Oct 2018, pp. 1--6.

\bibitem{PolarSpatiallyCoupled}
K.~{Wang}, W.~{Hou}, S.~{Lu}, P.~{Wu}, Y.~{Ueng}, and J.~{Cheng},
\newblock ``Improving polar codes by spatial coupling,''
\newblock in {\em International Symposium on Information Theory and Its
  Applications (ISITA)}, Singapore, Oct. 2018.

\bibitem{PolarInfoCoupled}
X.~{Wu}, L.~{Yang}, and J.~{Yuan},
\newblock ``Information coupled polar codes,''
\newblock in {\em IEEE International Symposium on Information Theory (ISIT)},
  Vail, CO, USA, June 2018.

\bibitem{Staircase_OFC}
C.~{Condo}, V.~{Bioglio}, and I.~{Land},
\newblock ``Staircase construction with non-systematic polar codes,''
\newblock in {\em Optical Fiber Communication Conference (OFC)}. 2020, p.
  Th1G.6, Optical Society of America.

\bibitem{TrifonovDesign}
P.~{Trifonov},
\newblock ``Efficient design and decoding of polar codes,''
\newblock {\em IEEE Transactions on Communications}, vol. 60, no. 11, pp.
  3221--3227, Nov. 2012.

\bibitem{leroux}
C.~Leroux, A.J. Raymond, G.~Sarkis, and W.J. Gross,
\newblock ``A semi-parallel successive-cancellation decoder for polar codes,''
\newblock {\em IEEE Transactions on Signal Processing}, vol. 61, no. 2, pp.
  289--299, Jan. 2013.

\bibitem{HDPC}
Ferdinand {Blomqvist},
\newblock ``{On Hard-Decision Decoding of Product Codes},''
\newblock {\em arXiv e-prints}, p. arXiv:2001.04715, Jan. 2020.

\bibitem{HDSC}
C.~{Hager} and H.~D. {Pfister},
\newblock ``Miscorrection-free decoding of staircase codes,''
\newblock in {\em 2017 European Conference on Optical Communication (ECOC)},
  2017, pp. 1--3.

\bibitem{SDPC}
R.~{Lucas}, M.~{Bossert}, and M.~{Breitbach},
\newblock ``On iterative soft-decision decoding of linear binary block codes
  and product codes,''
\newblock {\em IEEE Journal on Selected Areas in Communications}, vol. 16, no.
  2, pp. 276--296, 1998.

\bibitem{SDSC}
X.~{Dou}, M.~{Zhu}, J.~{Zhang}, and B.~{Bai},
\newblock ``Soft-decision based sliding-window decoding of staircase codes,''
\newblock in {\em International Symposium on Turbo Codes and Iterative
  Information Processing (ISTC)}, Dec. 2018.

\bibitem{Staircase2}
Lei~M. Zhang and Frank~R. Kschischang,
\newblock ``Staircase codes with 6\% to 33\% overhead,''
\newblock {\em J. Lightwave Technol.}, vol. 32, no. 10, pp. 1999--2002, May
  2014.

\bibitem{Bioglio:punct/short:WCNC2017}
F.~{Gabry} V.~{Bioglio} and I.~{Land},
\newblock ``Low-complexity puncturing and shortening of polar codes,''
\newblock in {\em IEEE Wireless Communications and Networking Conference
  (WCNC)}, San Francisco, CA, USA, Mar. 2017.

\bibitem{talSCL}
I.~{Tal} and A.~{Vardy},
\newblock ``List decoding of polar codes,''
\newblock {\em IEEE Transactions on Information Theory}, vol. 61, no. 5, pp.
  2213--2226, May 2015.

\bibitem{balatsoukas}
A.~Balatsoukas-Stimming, M.~Bastani~Parizi, and A.~Burg,
\newblock ``{LLR}-based successive cancellation list decoding of polar codes,''
\newblock {\em IEEE Transactions on Signal Processing}, vol. 63, no. 19, pp.
  5165--5179, Oct. 2015.

\bibitem{SCANfirst}
U.~U. {Fayyaz} and J.~R. {Barry},
\newblock ``Low-complexity soft-output decoding of polar codes,''
\newblock {\em IEEE Journal on Selected Areas in Communications}, vol. 32, no.
  5, pp. 958--966, May 2014.

\bibitem{SCANL}
C.~{Pillet}, C.~{Condo}, and V.~{Bioglio},
\newblock ``{SCAN} list decoding of polar codes,''
\newblock in {\em IEEE International Conference on Communications (ICC)},
  Dublin, Ireland, June 2020.

\bibitem{CA_SCL}
K.~{Niu} and K.~{Chen},
\newblock ``{CRC}-aided decoding of polar codes,''
\newblock {\em IEEE Communications Letters}, vol. 16, no. 10, pp. 1668--1671,
  October 2012.

\end{thebibliography}
